\documentclass[reprint,secnumarabic,amssymb, nobibnotes, aps, pre,groupedaddress,superscriptaddress]{revtex4-1}
\usepackage{graphicx}
\usepackage{dcolumn}
\usepackage{bm}
\usepackage{siunitx}
\usepackage[dvipsnames]{xcolor}
\usepackage{lipsum}
\usepackage{verbatim}
\usepackage{soul}
\usepackage{comment}
\usepackage{setspace}
\usepackage{amsmath}
\usepackage{amssymb}
\usepackage{pdfpages}
\usepackage{etoolbox} 

\makeatletter
\patchcmd{\@outputpage@head}{\@ifx{\LS@rot\@undefined}{}{\LS@rot}}{}{}{}
\makeatother

\newcommand{\beg}[1]{\begin{gather}#1\end{gather}}
\newcommand{\bea}[1]{\begin{align}#1\end{align}}
\newcommand{\vect}[1]{\boldsymbol{#1}}
\newcommand{\ddfrac}[2]{\frac{\displaystyle #1}{\displaystyle #2}}

\allowdisplaybreaks[1]

\begin{document}
\title{Weibel instability beyond bi-Maxwellian anisotropy}
\author{T. Silva}
\email{thales.silva@tecnico.ulisboa.pt}
\affiliation{GoLP/Instituto de Plasmas e Fus\~ao Nuclear, Instituto Superior T\'ecnico, Universidade de Lisboa, 1049-001 Lisbon, Portugal}
\author{B. Afeyan}
\email{bafeyan@gmail.com}
\affiliation{Polymath Research Inc., 94566 Pleasanton, CA, USA}
\author{L. O. Silva}%
\email{luis.silva@tecnico.ulisboa.pt}
\affiliation{GoLP/Instituto de Plasmas e Fus\~ao Nuclear, Instituto Superior T\'ecnico, Universidade de Lisboa, 1049-001 Lisbon, Portugal}
\date{\today}%

\begin{abstract}
	The shape of the anisotropic velocity distribution function, beyond the realm of strict Maxwellians can play a significant role in determining the evolution of the Weibel instability dictating the dynamics of self-generated magnetic fields. For non-Maxwellian distribution functions, we show that the direction of the maximum growth rate wavevector changes with shape. We investigate different laser-plasma interaction model distributions which show that their Weibel generated magnetic fields may require closer scrutiny beyond the second moment (temperature) anisotropy ratio characterization. 
\end{abstract}

\maketitle
\section{Introduction}
The anisotropy of typical plasma velocity distribution functions (VDFs) can fuel instabilities. The Weibel instability, in particular, is associated with some of the most striking astrophysical phenomena \cite{1999Gruzinov,1999Medvedev,2003Schlickeiser,2004Medvedev,2006Medvedev}, such as gamma-ray bursts, collisionless shocks, and magnetogenesis of the universe. It also plays a crucial role in laser-plasma interactions by directly coupling non-local heat transport, parametric instabilities, and spontaneous magnetic field generation. 

Weibel was the first to discover these unstable electromagnetic modes \cite{1959Weibel}, whose free energy source is typically attributed and characterized by parallel and perpendicular temperatures anisotropy assuming Maxwellians. Current density perturbations in the plasma interact with magnetic field perturbations; if the current density perturbation is in the hotter direction and the magnetic field wavevector in the colder, these fields act to reinforce the current perturbation, thus, generating an unstable feedback loop which leads to the spontaneous magnetic field growth. We challenge this attribution by showing that even for equal temperatures but different shapes in the parallel and perpendicular directions, Weibel unstable modes arise. We also show oblique Weibel modes can be dominant for widely different shaped VDFs. 

Due to its crucial role in high-energy-density plasmas and its connection with astrophysics, there is a renewed effort to measure the Weibel instability in the laboratory. Typically, counter-propagating plasma flows \cite{2015Huntington,2017Nerush} or laser-plasma interactions \cite{2019Zhang,2020Silva,2020Raj,2020Shukla,2020Zhang} drive the instability.  Frequently, laboratory and astrophysical plasmas operate in collisionless regimes, making significant deviations from Maxwellian distributions rather likely. Many astrophysical plasmas, such as the solar wind, are known for having hot non-thermal tails \cite{1968Vasyliunas}. Laser-plasma interactions are also known for generating hot \cite{1984Figueroa} or depleted tails \cite{1980Langdon,1988Matte}.
Non-Maxwellian distributions are rarely considered in the literature when performing kinetic analysis. Nevertheless, VDF details can be essential to estimate growth rates for the Weibel \cite{2010Lazar} and other kinetic instabilities \cite{1998Afeyan,2016ShaabanB,2016ShaabanA,2020Shaaban}.

In this work, we extend the Weibel instability analysis by exploring previously unidentified and unexpected features. Our results enable a deeper fundamental understanding of the instability mechanism. To go beyond standard theory, we consider a more generic spectrum of wavevectors and VDFs, emphasizing laser-plasma interaction generated VDFs. We use particle-in-cell simulations to confirm theoretical predictions and peer beyond the linear regime.

\section{A more general dispersion relation for the Weibel instability}
In the kinetic theory for the Weibel instability, one traditionally assumes Maxwellian VDFs with different temperatures in distinct directions, i.e.,
\begin{equation}
	f_0(v_x,v_y,v_z) = f_x^{\text{Max}}(v_x) f_y^{\text{Max}}(v_y) f_z^{\text{Max}}(v_z),
	\label{eq:bimax}    
\end{equation}
where $f_i^{\text{Max}}(v_i) = (2\pi T_i)^{-1/2} \mathrm{e}^{-v_i^2/2T_i}$ and $T_i$ is the temperature along the $i^{th}$ direction in units of $m_ec^2$ ($m_e$ is the electron mass and $c$ is the speed of light in vacuum). Considering immobile ions, the dispersion relation for electromagnetic modes with $\vect{k} = k_x \hat{x}$ and $\vect{k} = k_y \hat{y}$ is 
\begin{subequations}
	\begin{gather}
		k_x^2-\omega^2+1+\int\frac{ ~ \partial {f}_x/\partial {v}_x}{{v}_x-\omega/k_x}d{v}_x \int {v}_y^2 ~{f}_y d{v}_y  = 0,\label{eq:disp_simplex}\\
		k_y^2-\omega^2+1+\int {v}_x^2 ~{f}_x d{v}_x \int\frac{ ~ \partial {f}_y/\partial {v}_y}{{v}_y-\omega/k_y}d{v}_y = 0.\label{eq:disp_simpley}
	\end{gather}
	\label{eq:disp_simple}
\end{subequations}
Here, $v$ is given in units of $c$, $\omega$ in units of the electron plasma frequency $\omega_{p}=(4\pi n_e e^2/m_e)^{1/2}$ ($e$ and $n_e$ are the elementary charge and the plasma density), and $k$ in units of $\omega_{p}/c$. Assuming the VDF in Eq. \eqref{eq:bimax} and $T_x > T_y = T_z$, solutions of Eq. \eqref{eq:disp_simplex} are always damping modes for any $k_x$, whereas Eq. \eqref{eq:disp_simpley} has growing modes for $k_y^2 < (T_x/T_y-1)$ \cite{1973Krall}. Additionally, the highest growth rate is a solution of Eq. \eqref{eq:disp_simpley}.

Equation \eqref{eq:disp_simpley} dependence on the hotter direction comes as the temperature (we are generalizing the concept of temperature as the second moment of any distribution), so one could imagine replacing $f_x$ with another distribution would lead to identical results as for the same temperature Maxwellian. Although true for Eq. \eqref{eq:disp_simpley}, there is no guarantee that the maximum growth rate is a solution of Eq. \eqref{eq:disp_simpley} or that there are no growing solutions of Eq. \eqref{eq:disp_simplex}. The wavevector with the highest growth rate does not require to be aligned with either axis.

One derives a more general dispersion relation solving the Vlasov-Maxwell system using the method of characteristics. We consider an initially field-free plasma to understand how non-Maxwellian distributions affect the Weibel instability at a fundamental level. Although non-Maxwellian distributions often appear as the result of previous phenomena in the plasma that may induce fields, these fields typically seed certain unstable modes; the theory accurately predicts growth rates as long as these initial fields are small compared to the fields expected to grow in a field-free plasma. The solutions have the form $\vect{\mathcal{D}}\cdot \vect{E} = 0$ \cite{SM,2013Alexandrov}, where $\vect{E}$ is the electric field, $\mathcal{D}_{ij} = k_ik_j - k^2 \delta_{ij} + \omega^2 \varepsilon_{ij}$, $\delta_{ij}$ is the Kronecker delta, and $\varepsilon_{ij}$ is the dielectric tensor,
\begin{equation}
	\omega^2\varepsilon_{ij} = \left(\omega^2-1\right)\delta_{ij} + \sum_{\alpha=x,y}\int d^3 v \frac{k_\alpha v_iv_j \frac{\partial f_0}{\partial v_\alpha}}{\omega-k_xv_x-k_yv_y},
	\label{eq:dieletric}
\end{equation}
assuming now $\vect{k}=k_x \hat{x} + k_y \hat{y}$. This dispersion relation was studied previously in \cite{2014Vagin} for bi-Maxwellian VDFs. We focus on a specific class of VDFs, namely $f_0(v_x,v_y,v_z) = f_x (v_x) f_y^{\text{Max}}(v_y) f_z^{\text{Max}}(v_z)$, with $f_x (v_x) = f_x (-v_x)$ to assure current neutrality, and $T_y = T_z = T_\perp$. Under these assumptions, we note that the temperature tensor $T_{ij} \propto \int d^3v v_iv_jf_0$ is diagonal, which guarantees the phenomena observed, henceforth, are not due to a particular axes choice. This VDF type is relevant for high-energy-density physics where some laser-plasma interaction processes modify the VDF primarily in one direction. The transverse distribution $v_yv_z$ isotropy justifies the choice of $\vect{k}$, as it is possible to change coordinates such that $k_z = 0$. More generally, there will be a continuum of transverse wavevectors that grow. Consequently, the results will look more complicated, but they should be a superposition of the modes studied here.

Vlasov-Maxwell's system non-trivial solutions lead to the dispersion relation and require $\det(\mathcal{D}) = 0$, i.e., $\left(\mathcal{D}_{xx}\mathcal{D}_{yy}-\mathcal{D}_{xy}\mathcal{D}_{yx}\right)\mathcal{D}_{zz} = 0$ as we verified that $\mathcal{D}_{xz}=\mathcal{D}_{yz}=\mathcal{D}_{zx}=\mathcal{D}_{zy} \equiv 0$. Replacing the $f_0$ \textit{ansatz} in Eq. \eqref{eq:dieletric} and performing integration over $v_y$ and $v_z$, we obtain the relevant $\vect{\mathcal{D}}$ components as function of $f_x$, which are available at the Appendix. 
Our theory also reproduces the results from \cite{2007Tzoufras} for separable VDFs and pure wavevectors. Nevertheless, Weibel's instability general solutions must include mixed modes. Considering these modes are necessary to estimate maximum growth rates for some non-Maxwellian distributions and shed light on the instability mechanism.

The dispersion relation has two kinds of solutions,
\begin{subequations}
	\begin{gather}
		\mathcal{D}_{xx}\mathcal{D}_{yy}-\mathcal{D}_{xy}\mathcal{D}_{yx} = 0, \label{eq:dxy_0}\\
		\mathcal{D}_{zz} = 0.\label{eq:dzz_0}
	\end{gather}
	\label{eq:dxydzz_0}
\end{subequations}
Equation \eqref{eq:dzz_0} represents solutions with $\vect{E} = E_z \hat{z}$; due to Weibel's instability electromagnetic nature, Eq. \eqref{eq:dzz_0} unstable solutions result in the growth of magnetic field components $B_x$ and $B_y$. Analogously, unstable solutions of Eq. \eqref{eq:dxy_0} result in $B_z$ growth.

\section{The role of the shape of the velocity distribution function on Weibel unstable modes}
We use two example VDFs to highlight fundamental news aspects when using our theory. They are super-Gaussians and Maxwellians with hot tails, i.e.,
\begin{subequations}
	\begin{gather}
		f_x(v_x) = \mathcal{A}_m \mathrm{e}^{-\mathcal{B}_m\left|v_x\right|^m/T_x^{m/2}},\label{eq:dlm}\\
		f_x(v_x) = \frac{1 - \delta n}{(2\pi T_{\text{c}})^{1/2}}\mathrm{e}^{-v_x^2/2T_\text{c}}+\frac{\delta n}{(2\pi T_{\text{h}})^{1/2}}\mathrm{e}^{-v_x^2/2T_\text{h}},\label{eq:twotemp}
	\end{gather}
\end{subequations}
where $\mathcal{A}_m$ and $\mathcal{B}_m$ are such that $\int v_ x^2f_x dv_x = T_x$ and $\int f_x dv_x = 1$, $T_c$ and $T_h$ are the cold and hot population temperatures and $\delta n$ is the fraction of particles in the tail. The former distribution could result from ultrashort laser pulses field-ionizing the gas \cite{2019Zhang} or in the interaction between collisionless shocks \cite{2019Boella}; the latter can describe certain regimes of stimulated Raman scattering \cite{1980Estabrook}.

\begin{figure}[t]
	\centering
	\includegraphics[width=.99\linewidth]{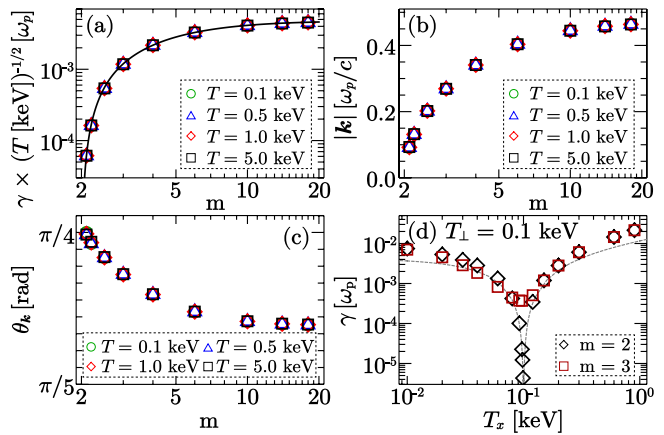}
	\caption{(a) Growth rate scaled by $\sqrt{T}$, (b) wavevector magnitude, and (c) wavevector angle for the highest growth rate mode as a function of the super-Gaussian exponent for different values of $T$ (symbols). In (a), the solid line is Eq. \eqref{eq:dlm_gr}. (d) Maximum growth rate for $m = 2$ and $m=3$ as a function of $T_x$ for fixed $T_\perp = \SI{0.1}{\kilo eV}$}.
	\label{fig:dlm_gr}
\end{figure}
We first explore super-Gaussian distributions with $T_x = T_\perp\equiv T$. Under this assumption, Equations \eqref{eq:disp_simplex} and \eqref{eq:disp_simpley} have only damping solutions, i.e., pure $k_x$ and $k_y$ modes are always stable. Unexpectedly, allowing mixed modes unlocks growing solutions of Eq. \eqref{eq:dxy_0} [Eq. \eqref{eq:dzz_0} has only damping modes]. Figures \ref{fig:dlm_gr}(a-c) characterize Eq. \eqref{eq:dxy_0} highest growth rate solution as a function of $m$. Figure \ref{fig:dlm_gr}(a) verifies that the growth rate $\gamma$ is proportional to $\sqrt{T}$ and increases monotonically with $m$. For the super-Gaussian distribution, we find an engineering formula for the growth rate as a function of $T$ and $m$ to be
\begin{equation}
	\frac{\gamma}{\omega_p} = 4.74\times 10^{-3}\left(\frac{T}{\SI{1}{\kilo eV}}\right)^{\frac{1}{2}}\tanh^4[0.884(m-2)^{2/5}], \label{eq:dlm_gr}
\end{equation}
which is the line in Fig. \ref{fig:dlm_gr}(a). Figures \ref{fig:dlm_gr}(b) and \ref{fig:dlm_gr}(c) display the wavevector magnitude $|\vect{k}| = (k_x^2+k_y^2)^{1/2}$ and angle $\theta_{\vect{k}} = \tan^{-1}(k_y/k_x)$, respectively. We note that $|\vect{k}|$ and $\theta_{\vect{k}}$ are independent of $T$. For increasing $m$, the wavenumber and angle vary from 0 and $\pi/4$ to the asymptotic values $\sim0.5~\omega_p/c$ and $\sim2\pi/9$, respectively.

\begin{figure}[t]
	\centering
	\includegraphics[width=.99\linewidth]{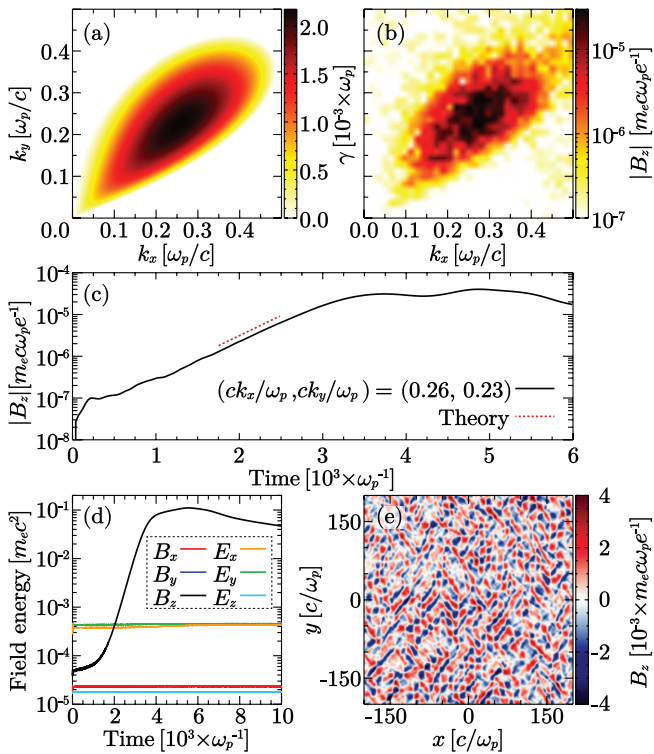}
	\caption{For a super-Gaussian with $m = 4$ and $T = \SI{1}{\kilo eV}$. (a) Theoretical growth rate as a function of the wavevector. (b) $|B_z|$ in Fourier space taken from simulations during the instability linear stage. (c) Comparison between the simulation and theory for the highest growth rate mode. (d) Electromagnetic field energy from the simulation. (e) Saturated $B_z$ in the configuration space.}
	\label{fig:dlm_sim}
\end{figure}
Figure \ref{fig:dlm_gr}(d) shows the maximum growth as a function of $T_x$ for $T_\perp = \SI{0.1}{\kilo eV}$. The dashed-gray line represents the standard Weibel theory approximate solution for low anisotropy ($A \equiv T_{\text{hot}}/T_{\text{cold}}-1 \approx 0$, where $T_{\text{hot/cold}}$ is the higher/lower of $T_x$ and $T_\perp$), which predicts no instability when $T_x = T_\perp$. We observe this trend for $m = 2$ (Maxwellian); for $m = 3$, the growth rate remains appreciable for all $T_x$. Thus, the growth rate could differ by several orders of magnitude depending on the VDF shape. When $A \approx 0$, it is fundamental to consider the VDF shape to predict the Weibel instability growth accurately. For large anisotropy, the maximum growth rate becomes identical to the Maxwellian case and, if $A\gg0$, follows the known $\gamma/\omega_p = (T_\text{hot}/m_ec^2)^{1/2}$ scaling \cite{1973Krall}.

To confirm our theoretical predictions, we performed two-dimensional particle-in-cell simulations using the OSIRIS framework \cite{2002Fonseca}. Simulation details are in  \cite{Sim01}. We compare theory and simulations in a case where $m=4$. Figure \ref{fig:dlm_sim}(a) shows the growth rate predicted by Eq. \eqref{eq:dxy_0} for a wide range of $\vect{k}$. Figure \ref{fig:dlm_sim}(a) is directly comparable with Fig. \ref{fig:dlm_sim}(b), the magnetic field in Fourier space in the simulation at $t = \SI{3000}{\omega_p^{-1}}$, i.e., during the instability linear stage, showing an excellent agreement between theory and simulation. The maximum growth rate in this example happens for the wavevector $(ck_x/\omega_p,ck_y/\omega_p)\approx(0.26,0.23)$ with $\gamma\approx\SI{0.002173}{\omega_p^{-1}}$; Figure \ref{fig:dlm_sim}(c) shows a direct comparison of this mode growth and the theoretical growth rate, also showing an excellent agreement. Figure \ref{fig:dlm_sim}(d) displays the energy evolution of all electromagnetic field components. Only the $B_z$ magnetic field component presents exponential growth, as predicted earlier for Eq. \eqref{eq:dxy_0} solutions; the electric field growth is negligible in comparison with the magnetic component, a known feature of the Weibel instability (cf., \cite{2017Cagas}). Although the modes observed are oblique, the electric field does not grow [Fig. \ref{fig:dlm_sim}(d)], and the magnetic component is consistent with a pure electromagnetic mode, making us confident that this is a manifestation of the Weibel rather than the oblique instability \cite{2009Bret,2010Bret}. Additionally, the VDF [Eq. \eqref{eq:dlm}] is not prone to the two-stream instability; hence there is no source for the electrostatic modes, which are an oblique instability component. For completeness, Figure \ref{fig:dlm_sim}(e) shows the saturated ($t=\SI{6000}{\omega_p^{-1}}$) magnetic field $B_z$ in the configuration space displaying the prevalence of oblique modes. The instability follows in excellent agreement with linear theory whereas the distribution function does not change appreciably; in this example, this lasts for around $\SI{2000}{\omega_p^{-1}}$. As the distribution tends to become isotropic, the growth rate decreases. In addition, the instability nonlinear saturation likely follows known arguments for the Weibel instability \cite{1972Davidson,1994Yang,2005Kato,2011Pokhotelov} and will not be explored further in this paper.

\begin{figure}[t]
	\centering
	\includegraphics[width=.99\linewidth]{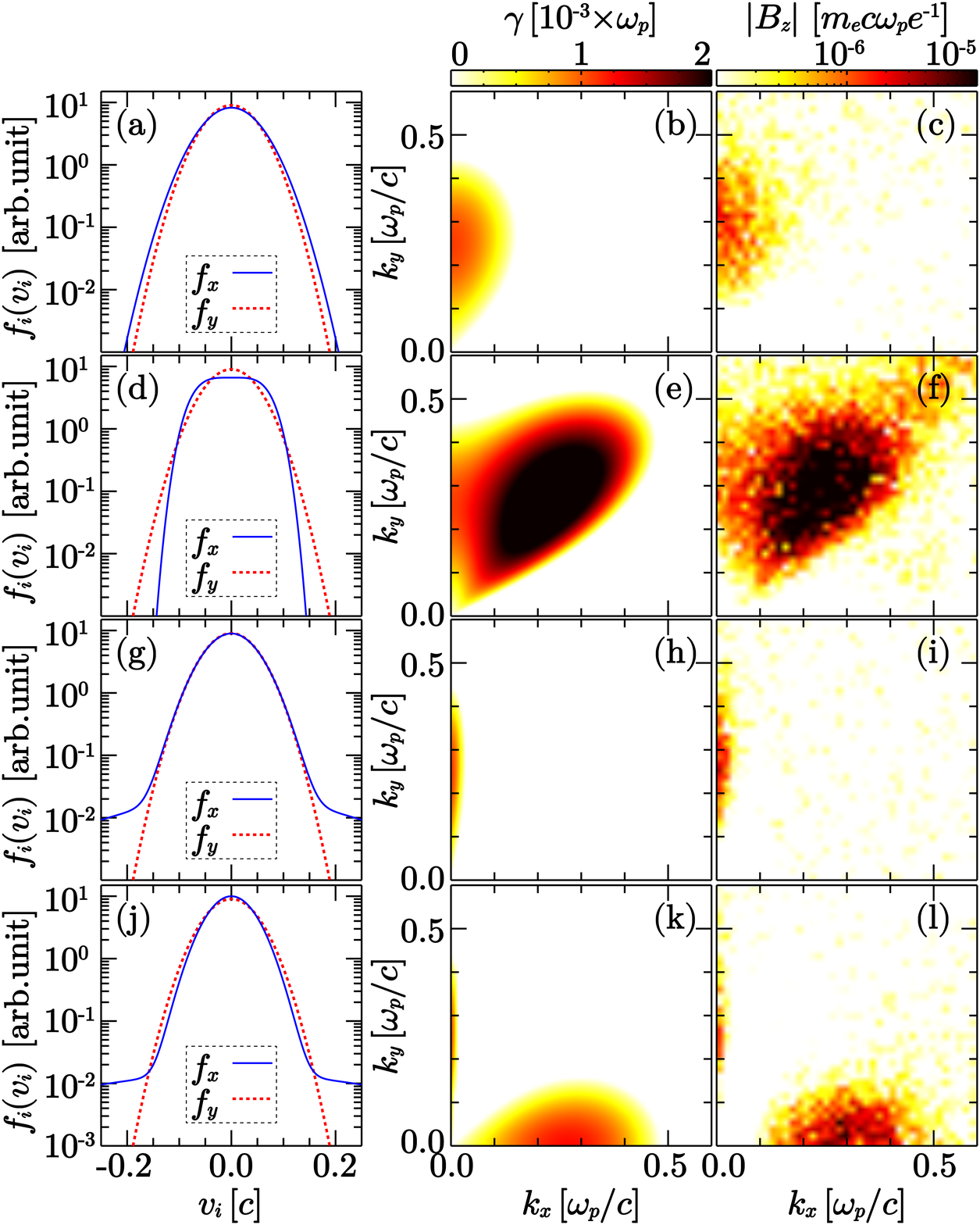}
	\caption{Initial VDF (left column), theoretical growth rate (center column), and $B_z$ from simulations (right column). (a-c) Maxwellian with $T_x = \SI{1.2}{\kilo eV}$; (d-f) super-Gaussian with $m = 4$ and $T_x = \SI{1.2}{\kilo eV}$; (g-i) hot tail with $T_c = \SI{1}{\kilo eV}$, $T_h = \SI{21}{\kilo eV}$, and $\delta n = 0.01$; and (j-l) hot tail with $T_c = \SI{0.8}{\kilo eV}$, $T_h = \SI{40.8}{\kilo eV}$, and $\delta n = 0.01$. $T_\perp = \SI{1}{\kilo eV}$ in all the examples.}
	\label{fig:t1.2kevcomp}
\end{figure}
The VDF shape plays a dominant role in determining the Weibel instability fundamental aspects, such as unstable wavevectors, their growth rates, and which VDF population part is the most relevant for the instability. To demonstrate those points, we compare four examples, all in which the effective temperature in the $x$-direction is $T_x=\int v_x^2 f_x dv_x = \SI{1.2}{\kilo eV}$, and $T_\perp = \SI{1}{\kilo eV}$. Figure \ref{fig:t1.2kevcomp} shows theory and particle-in-cell simulation results, with each row showing the initial VDFs $f_x$ and $f_y$ (left panel), the theoretical growth rate for a range of $\vect{k}$ (center panel), and $B_z$ in Fourier space at the instability linear stage taken from simulations (right panel).

Since the effective temperature $T_x$ is the same in all examples, theory predicts the same $\vect{k} = k_y \hat{y}$ modes. The remaining unstable solutions are different for each example. Figures \ref{fig:t1.2kevcomp}(a-c) show results for a Maxwellian VDF. In addition to the $\vect{k} = k_y \hat{y}$ expected modes, we notice that the unstable branch extends up to $k_x = 0.15~\omega_p/c$. Figures \ref{fig:t1.2kevcomp}(d-f) present results for a super-Gaussian ($m=4$) distribution, where we observe a wide range of oblique modes with higher growth rate than the $\vect{k} = k_y \hat{y}$. In Figs. \ref{fig:t1.2kevcomp}(g-i), we explore a hot tail distribution [Eq. \eqref{eq:twotemp}], where $T_c = T_\perp = \SI{1}{\kilo eV}$, $\delta n = 1\%$, and $T_h = \SI{21}{\kilo eV}$. We observe a narrow, unstable branch with $\vect{k} \approx k_y\hat{y}$, confirmed in simulations. Figures \ref{fig:t1.2kevcomp}(j-l) show a second hot tail distribution example, with $T_c = \SI{0.8}{\kilo eV}$, $\delta n = 0.01$, and $T_h=\SI{40.8}{\kilo eV}$. We also observe a narrow branch with $\vect{k} \approx k_y\hat{y}$, but this time there is a larger branch with modes $\vect{k} \approx k_x\hat{x}$ that extends up to $k_y = 0.15~\omega_p/c$. The latter branch is commensurate with the solution when $\delta n = 0$, i.e., a bi-Maxwellian with $T_x = \SI{0.8}{\kilo eV}$ and $T_\perp = \SI{1.0}{\kilo eV}$. We confirmed the two separate branches presence in simulations [Fig. \ref{fig:t1.2kevcomp}(l)]. The mode with the highest growth rate is in the hotter direction ($\vect{k} = k_x\hat{x}$), demonstrating it does not necessarily lie along the colder direction for non-Maxwellian VDFs. We emphasize that the presence of a larger number of unstable modes has a direct consequence in the generated magnetic field strength; the saturated magnetic field amplitude is about one order of magnitude higher for the Maxwellian case [Fig. \ref{fig:t1.2kevcomp}(a)] when compared with the hot tail [Fig. \ref{fig:t1.2kevcomp}(g)].

\section{A proposed generalized metric for predicting the Weibel unstable modes}
The field-free Vlasov equation has infinite solutions of the kind $f_0=f_0(v^2)$, and anisotropic (but current neutral) VDFs evolve in ways that generate magnetic fields in the plasma through the Weibel instability. The temperature tensor $T_{ij} \propto \int d^3v v_iv_jf_0$ is already diagonal in the $v_xv_y$ coordinates for the examples in Fig. \ref{fig:t1.2kevcomp}. Thus, the traditional temperature measurement cannot explain the presence of oblique modes and multiple branches. We conjecture that the instability will rise from any anisotropy in the VDF, and the generated magnetic field wavevector angle is perpendicular to the maxima VDF spread directions. To test the conjecture, we define the quantity
\begin{equation}
	\varepsilon_\theta = \ddfrac{\int_{-\infty}^{\infty} v^2 f_x(v\cos\theta) f_y(v\sin\theta) dv}{\int_{-\infty}^{\infty} f_x(v\cos\theta) f_y(v\sin\theta) dv},
	\label{eq:ene_theta}
\end{equation}
the VDF dispersion along the direction $\theta$ about $v_x$. This quantity mainly differs from the temperature tensor because integration is made along one direction and not all velocity space.

\begin{figure}[t]
	\centering
	\includegraphics[width=.99\linewidth]{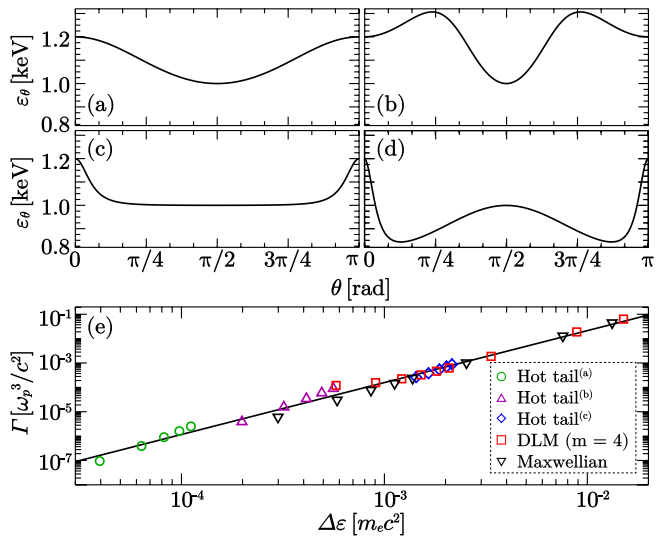}
	\caption{(a-d) $\varepsilon_\theta$ given in Eq. \eqref{eq:ene_theta} for the same examples in Fig. \ref{fig:t1.2kevcomp}. (e) Relation between $\Delta\varepsilon$ and $\Gamma$ for several examples. Hot tail$^\text{(a)}$: $T_c = \SI{1.0}{\kilo eV}$ and $\delta n = 0.01$, with $T_h$ varying from \SIrange{3}{11}{\kilo eV}. Hot tail$^\text{(b)}$: $T_c = \SI{1.0}{\kilo eV}$ and $\delta n = 0.05$, with $T_h$ varying from \SIrange{3}{11}{\kilo eV}. Hot tail$^\text{(c)}$: $T_c = \SI{1.5}{\kilo eV}$ and $\delta n = 0.05$, with $T_h$ varying from \SIrange{2}{20}{\kilo eV}. Super-Gaussian with $m = 4$ and $T_x =$ \SIrange{1}{10}{\kilo eV}. Maxwellian with $T_x =$ \SIrange{1}{10}{\kilo eV}. $T_\perp = \SI{1}{\kilo eV}$ in all the examples. The solid line represents Eq. \eqref{eq:conj}.}
	\label{fig:conject}
\end{figure}

Figure \ref{fig:conject}(a-d) shows $\varepsilon_\theta$ as a function of $\theta$ for the Fig. \ref{fig:t1.2kevcomp} examples in the order of appearance. We notice $\varepsilon_0 = \SI{1.2}{\kilo eV}$ and $\varepsilon_{\pi/2} =  \SI{1.0}{\kilo eV} $ agree with the diagonal temperature tensor components for all the examples; otherwise, the behavior as a function of $\theta$ varies significantly. The maxima of $\epsilon_\theta$ indicate the presence of unstable branches. Panels (a) and (c) for $f_x$ Maxwellian and hot tail with $T_c = T_y$, the maximum of $\varepsilon_\theta$ for $\theta = 0$ points to unstable branches at $\theta = \pi/2$ as observed in Figs. \ref{fig:t1.2kevcomp}(b) and \ref{fig:t1.2kevcomp}(h). Additionally, the unstable branch size seems to be related to the excess area above the minimum value of  $\varepsilon_{\theta = \pi/2} = \SI{1.0}{\kilo eV}$. In Fig. \ref{fig:conject}(b), for the super-Gaussian example, the maximum at the oblique angle $\theta\approx \SI{0.76\pi}{rad}$ implies the unstable branch at $\theta\approx \SI{0.26\pi}{rad}$, in good agreement with the theoretical maximum growth rate $\tan^{-1}(k_y/k_x) \approx \SI{0.27\pi}{rad}$. An analogous calculation for the example of Fig. \ref{fig:dlm_sim} gives the unstable branch at $\theta\approx \SI{0.21\pi}{rad}$, whereas the theory predicts $\tan^{-1}(k_y/k_x) \approx \SI{0.23\pi}{rad}$. Finally, the different maxima at $\theta = 0$ and $\theta = \pi/2$ in Fig. \ref{fig:conject}(d) can explain the two distinct branches in Fig. \ref{fig:t1.2kevcomp}(k).

To demonstrate that the excess area above the lowest value of $\varepsilon_\theta$ is related to the unstable branch size and growth rates, we define the quantity $\Gamma = \int_{\gamma>0} \gamma\left(k_x,k_y\right) dk_x dk_y$, the Fourier space area with unstable solutions ($\gamma >0$) weighted by the growth rate of each mode. We compare with the quantity $\Delta\varepsilon = \int_0^\pi \left(\varepsilon_\theta - \varepsilon_\theta^{\text{min}}\right)d\theta$,
where $\varepsilon_\theta^{\text{min}}$ is the minimum value of $\varepsilon_\theta$. Note that $\Delta\varepsilon = 0$ if $f_0$ is isotropic (Vlasov's equation stationary solution) and there is no instability.

Figure \ref{fig:conject}(e) compares $\Gamma$ and $\Delta\varepsilon$ for several $f_x$ and parameters with $f_y$ being a Maxwellian with $T_\perp = \SI{1}{\kilo eV}$. The relation between $\Gamma$ and $\Delta\varepsilon$ is well described by a power law
\begin{equation}
	\Gamma = 390 \frac{\omega_p^3}{c^2} \times \left(\frac{\varepsilon_\theta^{\text{min}}}{\SI{1}{\kilo eV}}\right)^{-1.75} \times \left(\frac{\Delta\varepsilon}{m_e c^2}\right)^{2.13}.
	\label{eq:conj}
\end{equation}
We verified Eq. \eqref{eq:conj} for other values of $\varepsilon_\theta^{\text{min}}$ not shown in Fig. \ref{fig:conject}(e). Although we do not have a proof that this is the most general mechanism, it assuredly leads to a better understanding of the Weibel instability. Previous metrics rely on the temperatures or the anisotropy parameter \cite{2017Schoeffler} and can only reliably determine the maximum growth rate for bi-Maxwellian VDFs. Even for these VDFs, there is a range of unstable oblique wavevectors that significantly contribute to the magnetic field generated as demonstrated in Fig.  \ref{fig:t1.2kevcomp}(b). In addition, the metric $\varepsilon_\theta$ predicts instability even when $f_x = f_y$ if $f_0 \ne f_0(v^2)$, which was confirmed in simulations with $f_x$ and $f_y$ being super-Gaussians ($m=4$) with $T_x = T_y = \SI{1}{\kilo eV}$.

\section{Conclusions}
In this work, we have studied the Weibel instability for non-Maxwellian VDFs and allowing oblique wavevectors. We have shown that the VDF shape plays a significant role to determine the maximum growth rate and unstable modes. We have derived an empirical formula for the maximum growth rate when the VDF is super-Gaussian along one direction and the temperature anisotropy is small. We have also shown that a better measurement for the Weibel instability is based on the VDF spread excess along a particular direction rather than the temperature. Such a quantity leads to a better grasp of the full-range Weibel unstable modes than by only looking at the temperatures. We have explored examples typical from laser-plasma interactions, thus, showing that Weibel fields generated in those scenarios may need to consider the VDF shape to correctly characterize the magnetic fields observed.

\begin{acknowledgments}
	TS is thankful for discussions with Dr. Kevin Schoeffler. We gratefully acknowledge computing time on the GCS Supercomputer SuperMUC (Germany); and PRACE for awarding us access to MareNostrum at BSC (Spain). This work was supported by the European Research Council through the \mbox{InPairs} project Grant Agreement No. 695088, FCT (Portugal) grant PTDC/FIS-PLA/2940/2014 and UID/FIS/50010/2023. The work of BA was supported by a grant from the DOE FES-NNSA Joint program in HEDLP DE-SC 0018283. 
\end{acknowledgments}
\appendix*
\section{$\mathcal{D}_{ij}$ tensor components}
\label{app:1}
Below, we present the $\mathcal{D}_{ij}$ tensor components. Given a $f_x$ distribution, we calculate the integrals using QUADPACK and then solve the dispersion relation [Eq. \eqref{eq:dxydzz_0}] using the bisection method. The tensor components are
\bea{
&\mathcal{D}_{xx} = \omega^2 - k_y^2 + \frac{T_x}{T_y} -1 - \frac{k_x}{k_y} \frac{1}{\sqrt{2T_y}} \int dv_x v_x^2 f^\prime_x Z(\xi)  \nonumber\\
&\quad\quad\quad\quad\quad\quad\quad\quad\quad\quad\quad~\quad\quad\quad+ \frac{1}{T_y} \int dv_x v_x^2 f_x \xi Z(\xi), \nonumber \\
&\mathcal{D}_{xy} = \mathcal{D}_{yx} = k_xk_y + \frac{k_x}{k_y} -\frac{k_x}{k_y} \int dv_x v_x f^\prime_x \xi Z(\xi) \nonumber\\
&\quad\quad\quad\quad\quad\quad\quad\quad\quad\quad~~+ \sqrt{\frac{2}{T_y}} \int dv_x v_x f_x \xi[1+\xi Z(\xi)], \nonumber 
}
\bea{
&\mathcal{D}_{yy} = \omega^2 - k_x^2 - \frac{k_x}{k_y}\sqrt{2T_y} \int dv_x f^\prime_x \xi[1+\xi Z(\xi)] \nonumber\\
&\quad\quad\quad\quad\quad\quad\quad\quad\quad\quad\quad\quad\quad+ 2 \int dv_x f_x \xi^2[1+\xi Z(\xi)], \nonumber \\
&\mathcal{D}_{zz} = \omega^2 - k_x^2 - k_y^2 + \frac{T_z}{T_y} -1 - \frac{k_x}{k_y} \frac{T_z}{\sqrt{2T_y}} \int dv_x f^\prime_x Z(\xi) \nonumber\\
&\quad\quad\quad\quad\quad\quad\quad\quad\quad\quad\quad\quad\quad\quad\quad~+ \frac{T_z}{T_y} \int dv_x f_x \xi Z(\xi),\nonumber
}
where 
\beg{
\xi = \frac{\omega-k_xv_x}{\sqrt{2T_y}k_y},~~\text{and}~~ Z(\xi) = \pi^{-1/2} \int \frac{\mathrm{e}^{-v^2}}{v-\xi} dv. \nonumber
}

\reversemarginpar
\clearpage
\onecolumngrid
{\setstretch{1.25}
\setcounter{figure}{0}
\renewcommand{\thefigure}{\Roman{figure}}

\begin{thebibliography}{39}
\bibitem [{Sim()}]{Sim01}%
  \BibitemOpen
  \href@noop {} {}\bibinfo {howpublished} {Simulations were performed with
  periodic boundaries. The box size is
  $400~c/\omega_{pe}\times400~c/\omega_{pe}$ divided into a grid with square
  cells with edges $0.1~c/\omega_{pe}$ long. Each cell starts with 15625
  particles.}\BibitemShut {Stop}%
\end{thebibliography}
\end{document}